# Solving an old puzzle: fine structure of diffraction spots from an azo-polymer surface relief grating

Joachim Jelken, Carsten Henkel[*], Svetlana Santer

*Institute of Physics and Astronomy, University of Potsdam, 14476 Potsdam, Germany*

AUTHOR EMAIL ADDRESS: henkel@uni-potsdam.de

ABSTRACT

We report on the experimental and theoretical interpretation of the diffraction of a probe beam during inscription of a surface relief grating with an interference pattern into a photo-responsive polymer film. For this we developed a set-up allowing for the simultaneous recording of the diffraction efficiency (DE), the fine structure of the diffraction spot and the topographical changes, *in situ* and in real time while the film is irradiated. The time dependence of the DE, as the surface relief deepens, follows a Bessel function exhibiting maxima and minima. The size of the probe beam relative to the inscribed grating (i.e., to the size of the writing beams) matters and has to be considered for the interpretation of the DE signal. It is also at the origin of a fine structure within the diffraction spot where ring-shaped features appear once an irradiation time corresponding to the first maximum of the DE has been exceeded.

---

[*] Corresponding author: henkel@uni-potsdam.de

**Introduction**

Azobenzene containing polymer thin films can develop significant opto-mechanical stress under irradiation with spatially modulated light[1,2] which results in the macroscopic deformation of the surface. The deformation ranges from elongation or contraction in one direction during irradiation with linear polarized light[3] to a fine vortex-like topographical response to a complex shape of the inscribing beam.[4,5,6] Special relief structures can be generated by applying illumination with an interference pattern, where the incoming light shows a periodic variation of the magnitude or local orientation of the electrical field vector.[7,8] The dynamical responsive of these polymeric materials is such that the topography deforms on time scales of seconds to minutes, mimicking the optical interference pattern, and establishes a sinusoidal profile, so-called surface relief grating (SRG), whose period is in most cases equal to the optical periodicity.[9,10,11,12,13] It was demonstrated that such topographical gratings can have periods as small as 125nm for a certain polarization pattern (SP configuration: one writing beam s-polarized, the other one p-polarized), while the height of the pattern can be as large as ~90% of the total thickness of the polymer film (in the RL configuration: one beam with right-handed circular polarization, the other beam left-handed).[14,15] The process takes place in air without additional solvent or temperature softening, under rather low intensity ($I \sim 50mW/cm^2$). This is remarkable because the polymer is in a solid (glassy) state and its mechanical modulus is several hundred of MPa. A possible explanation of such a deformation is the orientation model proposed by Saphiannikova et al.[16,17] In this model the SRG formation is considered as a multiscale chain of several processes starting from the small scale motion of azobenzene molecules under cyclic *trans-cis-trans* isomerization, which causes the local alignment of azobenzene groups perpendicularly to the electrical field vector, followed by re-orientation of the polymer backbones they are connected to.[18,19] The process ends up generating strong internal, anisotropic stress and the subsequent macroscopic opto-mechanical deformation of the film, as manifested in the SRG.[20,21,22,23,24,25] During this process the formation of two gratings can be distinguished: a birefringence grating in the bulk due to the local alignment of azobenzene chromophores, and a surface grating as a result of topographical deformation. Experimentally the SRG formation process can be probed by focusing a red laser beam on the polymer film and measuring the intensity of the diffracted light as a function of irradiation time. (At the probe wavelength, the material absorption is relatively weak.) The changes in the diffraction efficiency (DE) signal follow a $n$'th order Bessel function and can be explained using a time-dependent model based on the Raman-Nath diffraction theory.[26] It has been reported, indeed, that the diffraction efficiency is not monotonously increasing as a function of irradiation time, but can drop in the course of grating development. [27,28,29,30,31,32] However, the DE signal contains information about bulk and surface gratings and requires a quite involved de-convolution model in order to separate the contributions of both gratings to the time-dependent DE signal.[33,34,35,36]

Recently we have proposed a set-up where one can directly separate birefringence and surface gratings by measuring the development of DE signal and topography simultaneously and *in-situ*, i.e. during irradiation.[37] The set-up consists of three parts (see **Scheme 1**): (1) a two-beam interference system for generating a well-defined intensity or polarization interference pattern; (2) a probe laser with a set of photodiodes for recording the diffraction efficiency, and (3) an atomic force microscope (AFM) for acquiring the change in surface topography during irradiation. The DE and AFM data are taken simultaneously over time scales of seconds to minutes, while the SRG grows in amplitude in the presence of the interference pattern. In this paper, complementing the above described set-up, we report on data taken with an additional component that acquires the spatial intensity profile of the diffraction spot. Moreover, we explain a maximum and subsequent decrease in the DE that appears during the development of the SRG when its amplitude typically exceeds around 100nm. The explanation is supported by calculations using a "reflecting phase screen" model in the Raman-Nath approximation,[26] and is based on the fact that the profile of the writing beam has a Gaussian shape. The corresponding modulation depth of the SRG changes over the inscribed area (see **Figure 4b** below): the grating has a maximal amplitude in the center of the writing spot which drops radially to periphery. The size of the probe beam that records the SRG growth, relative to the inscribed area, determines the conditions where the DE signal peaks occur. Using a wide probe beam, which matches the inscribed grating area in size, one finds a fine structure in the spatial profile of the diffraction spot, which has been reported earlier by other groups.[38] The spatial profile of the diffraction spot changes as a function of the SRG height starting from a Gaussian to a hollow ring ("donut" like) and finally to a ring structure with a bright center. The hollow ring appears at the moment where the total DE signal decreases, while the ring with the bright center sets in when the DE signal starts to increase again. On the example of two different polymers, at fixed size of the writing and probe beams, we show that the decrease in the DE signal and the variation of the fine structure of the beam profile of diffraction spot starts at the same SRG height. For a very small probe beam diameter the diffraction efficiency follows the prediction for a homogeneous grating of infinite area (given by squares of ordinary Bessel functions).

**Experimental Part**

**Materials and Methods**

*Poly[1-[4-(3-carboxy-4-hydroxyphenylazo)benzenesulfonamido]-1,2-ethanediyl, sodium salt]* (Pazo) and *Poly[(methyl methacrylate)-co-(Disperse Red 1 acrylate)]* (poly(MMA-co-DR1A)) are purchased from Sigma-Aldrich. The Pazo polymer solution is prepared by dissolving 170mg Pazo in 1ml solution containing a mixture of 95% methoxyethanol and 5% ethylene glycol. The poly(MMA-co-DR1A) polymer is dissolved in chloroform to get a concentration of 60mg/ml.

*Sample preparation*. The polymer films are prepared by spin casting at 3000 rpm for one minute of 100μl of the polymer solution on a thin glass slide. The film thickness is measured from the cross-sectional profile of an atomic force microscope (AFM) micrographs along a scratch within the polymer film.

**Methods**

*Home-made set-up for studying SRG formation in-situ.* The set-up consists of three parts: (1) two beam interference lithography, (2) atomic force microscopy (AFM) and (3) diffraction efficiency (DE) measurement (**Scheme 1**). The two beam interference lithography with a continuous wave diode pumped solid state laser (Cobolt Calypso, $\lambda$=491nm) allows generating well-defined spatiotemporal interference patterns by changing the polarization of two interfering beams in a controlled way. In this part of the set-up the laser beam is spatially expanded and then collimated with a pair of focusing and collimating lenses and a pinhole (**Scheme 1**). The beam diameter is set to 4mm and the total intensity is varied between 100mW/cm$^2$ and 200mW/cm$^2$. Additionally, a 50:50 beam splitter is added in order to separate the initially single beam into two beams of the same intensity. These two beams pass through a set of wave plates and polarizers allowing independent control of intensity and polarization. For instance, adding a half-wave plate to each of the beam paths of the interference set-up (H$_3$, H$_4$), one with an angle of +22.5° and the second with an angle of −22.5° with respect to the optical axis, results in the ±45° interference pattern (IP). This is a polarization interference pattern with constant intensity in the sample plane, but spatially varying distribution of polarization.

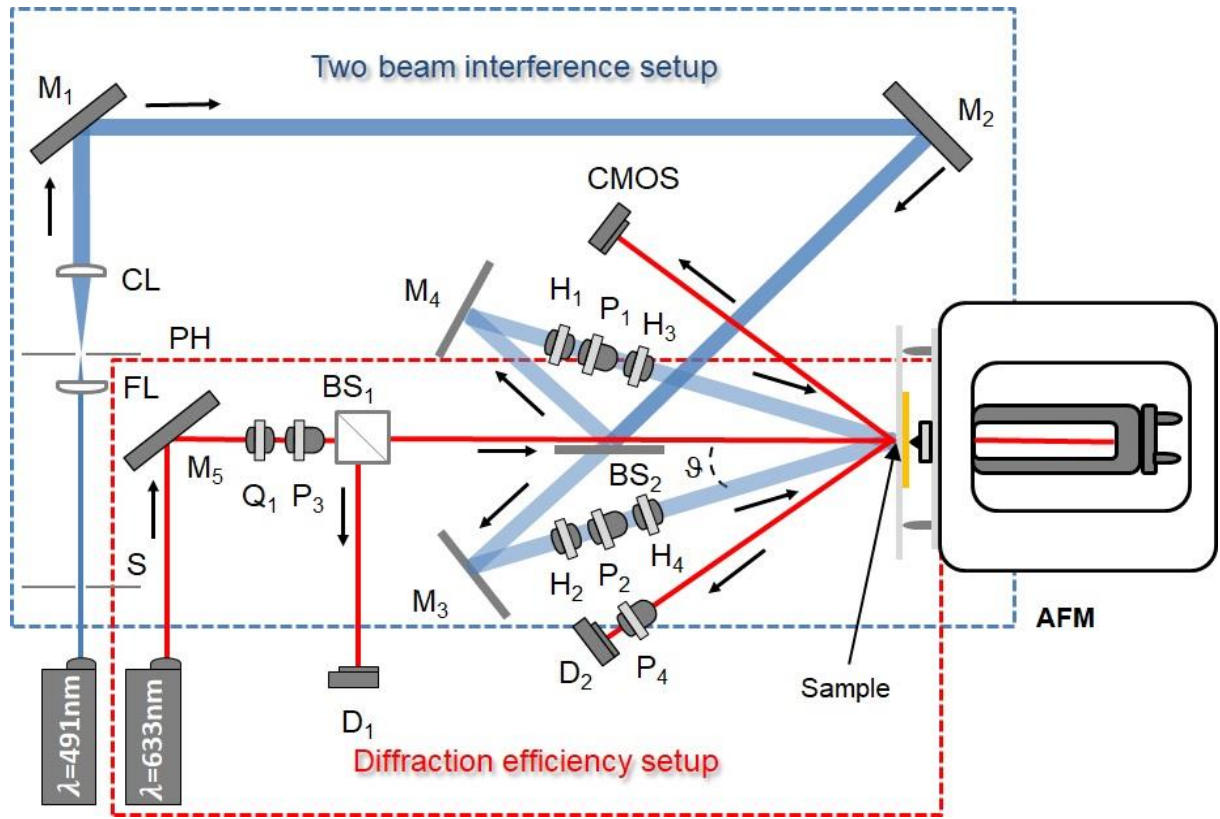

**Scheme 1.** Scheme of the experimental set-up consisting of three parts: (1) a two-beam interference part for generating the interference pattern (blue laser beams), (2) a diffraction efficiency (DE) set-up (red laser beams) enabling the collection of *in-situ* information about the optical grating (periodic refractive index and surface relief), and (3) an atomic force microscope (AFM) for *in-situ* recording of the surface morphology (during irradiation). (S: shutter, M: mirror, D: detector, P: polarizer, H: half-wave-plate, Q: quarter-wave-plate, BS: beam splitter, CL: collimating lens, FL: focusing lens, PH: pin hole, CMOS: camera.)

The second part of the home-made set-up is an atomic force microscope (AFM, PicoScan (Molecular Imaging) AFM operating in intermittent contact mode) enabling measurements of the polymer topography changes *in-situ* and in real time, i.e. under varying irradiation conditions. The scan-speed of the AFM is set to 1Hz with a scan-area of 10x10μm and a resolution of 512x512 pixel. Commercial tips (Nanoworld-Point probe) with a resonance frequency of 130 kHz, and a spring constant of 15 N/m are used. The sample is oriented with the polymer surface pointing towards the AFM tip, such that irradiation is "from below", i.e., through the glass substrate (**Scheme 1**). The SRG amplitude is determined from the plot of the surface profile of the AFM scan by measuring the difference in height between topography maximum and minimum. A plot of this value as a function of time gives the growth kinetic of the SRG. To obtain at the same time information about the alignment of the azobenzene side chains in the polymer film, a red probe laser beam (Uniphase, HeNe, 633nm) with a tunable beam diameter from 1mm to 3mm ($I = 30mW/cm^2$) is integrated into the set-up. The wavelength of 633nm is outside the absorption band of both polymers studied in this work and its intensity is weak enough not to affect the polymer film, as proved by AFM measurements. To calibrate the DE, a beam splitter with a ratio (T:R = 90:10) is used in the DE set-up, such that 90% of the light arrives on the sample (intensity $I_0$) and 10% on a photodiode. The signal of this photodiode is recorded during the whole measurement for controlling the stability of the probe beam during the experiment. The diffraction efficiency is defined as the ratio of the intensity of diffraction order ($I_n$) and the intensity of the incoming light ($I_0$): $\eta_n = \frac{I_n}{I_0}$, where $I_0$ is 90% of the total intensity of the probe beam. The DE set-up additionally includes a quarter-wave plate converting the polarization of the probe beam from linear to circular. The polarization state of the probe beam can be set by adding a polarizer afterwards. The probe beam is P-polarized for all measurements discussed in this paper. The detector D2 measures the P- polarized component of the 1$^{st}$ order DE signal.

The three different set-ups: two beam interference, AFM and DE acquisition are controlled and operated with a software (Profilab-Expert, Abacom) specifically designed in the laboratory to record signals of the photodiodes, control the irradiation shutter and to synchronize DE-set-up with the AFM. A computer-generated signal regulates the irradiation with the help of an AD/DA converter (Kolter Electronic, PCI-AD12N-DAC2), which also records the signals of the photodiodes. This irradiation signal is sent to the diffraction efficiency set-up as well as to the AFM by recording the signal with the aux-input of the AFM controller, which enables to synchronize the different components.

The set-up is aligned in such a way that the AFM probe is placed in the center of the two interfering beams, using as a reference spot the red laser of the AFM optical lever focused on the cantilever. Afterwards the probe beam is aligned to the center of the IP and AFM probe.

*Silicon-detectors* (Thorlabs DET 100A/M) are used in the diffraction efficiency (DE) set-up to measure the intensity of the diffracted probe beam. A 600nm longpass filter is placed in front of each photodiode in order to be only sensitive to the probe beam. The intensity of the diffracted light is recorded every 200ms.

The change in the beam profile of the first order DE is recorded with a CMOS Camera (Thorlabs DCC1545M) with a sensor size of 6.7 x 5.3 mm and resolution of 1280 x 1024 pixel (pixel size 5.2 µm). For these measurements the photodiode is replaced by the CMOS camera; alternatively, the camera can be placed in the direction of the -1 diffraction order (see **Scheme 1**).

All experiments are carried out under yellow light in the laboratory (to avoid undesirable photo-isomerization) and under ambient conditions, *i.e.*, at room temperature with a relative humidity of 55%. The whole set-up is covered with a non-transparent encapsulation in order to avoid any influence of the environment on the measurement (parasitic light, air circulation and vibration).

**Results and Discussion**

**Figure 1** shows a typical curve recorded during irradiation of the azobenzene containing polymer film (Pazo) with the ±45° interference pattern (IP). The SRG height (black dots in **Figure 1**) increases continuously, while the DE signal (1$^{st.}$ order, red curve in **Figure 1a**) has a Bessel function shape where several maxima and minima appear. Indeed, the DE signal first increases within 10 minutes of irradiation followed by a decrease when the SRG height reaches 100±10 nm. In the course of irradiation during 6 hours two maxima of the DE signal (at SRG height of 100±10 nm and 300±10 nm) and two minima (at 200±10 nm and 400±10 nm) develop (**Figures 1c, d**). Similar behavior is observed for other polymers, here illustrated with poly(MMA-co-DR1A) (**Figure S1b, Supporting Information**). The extremal points of the DE signal appear at the same positions relative to the SRG height, although the SRG kinetic growth is much faster for the poly(MMA-co-DR1A) (**Figure S1a, Supporting Information**).

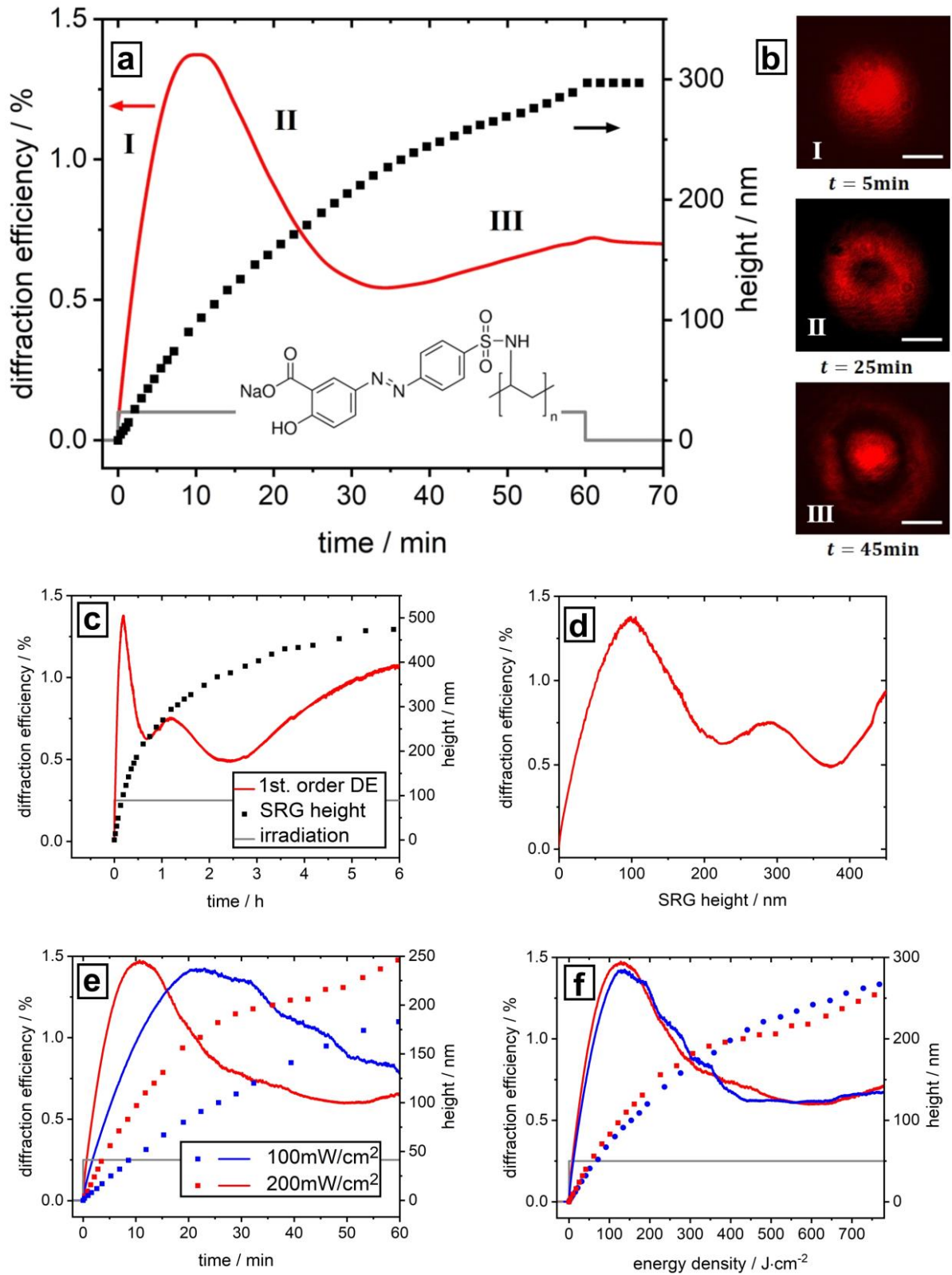

**Figure 1.** (a, c) *In-situ* recorded SRG height and diffraction efficiency (DE) under irradiation of a Pazo film with a $\pm 45°$ interference pattern (IP) of λ=491nm light ($I = 200 \text{mW/cm}^2, \Lambda = 2\mu\text{m}$ grating period, thickness $d = 1\mu\text{m}$); (a) first hour of irradiation; (c) irradiation during six hours. Note the non-monotonous increase and decrease in the $1^{st}$ order DE (red curve), while the SRG height increases continuously (black dots). The inset depicts the chemical structure of the Pazo polymer. (b) Spatial profile of the 1st order diffraction spot for three irradiation times (camera images, scale bar 1mm). The shape is Gaussian, until the maximum DE signal (SRG height of 100±10 nm, micrograph marked by I) is reached. When the DE starts to decrease, there is a noticeable change in the beam profile (micrographs marked by II and III). The intensity in the center of the diffraction spot decreases, resulting in a ring-shaped distribution (II). Further irradiation changes the spot shape to a ring structure with a bright center (III). (d) Parametric representation of the data in (c): DE (optical data) as a function of SRG height (AFM data). (e) *In-situ* recorded SRG height (dots) and 1st order DE (solid line) for two irradiation intensities: 200mW/cm$^2$ (red) and 100mW/cm$^2$ (blue). (f) Plot of the data of (e) as a function of fluence (intensity multiplied by time).

The DE signal typically measured is the integrated value over the whole detector area, i.e. the entire diffraction spot. The analysis of spot profile in the 1st. order reveals, however, a time-dependent intensity modulation resembling "donut" structure. As long as the DE signal grows (first 10 minutes of irradiation), the beam profile has a Gaussian shape (micrograph I in **Figure 1b**). At maximum DE signal (the SRG height is 100±10 nm), the intensity in the center of the laser spot (micrograph II in **Figure 1b**) starts to decrease (appearance of the "donut"). Further irradiation with an interference pattern will modify a spot profile further into a ring structure with a bright center (after 45 min, micrograph III in **Figure 1b**). The time evolution of diffracton spot profile is shown in a video (see **Supporting Information, Figure S2**). The DE slightly decreases when the pump beam is switched off (after 60min of irradiation), while the SRG amplitude and the spatial profile of the 1st order diffraction spot do not change. This indicates directly the relaxation of the birefringence grating. When the irradiation intensity is reduced in two times (from 200mW/cm$^2$ to 100mW/cm$^2$), a similar behavior of the DE signal is observed (blue curve in **Figure 1e**), but at longer irradiation time where the SRG has reached to the height of 100±10 nm. This behavior scales with the energy fed into the system, as illustrated by the collapse of the DE and SRG data when plotted as a function of the product of intensity and irradiation time (insert in **Figure 1f**).

At constant irradiation parameters, the kinetic of the SRG growth and the maximal SRG height depends on the thickness of the polymer film. **Figure 2** shows a comparison of the $1^{st.}$ order DE and SRG height for different thicknesses of the Pazo polymer film and a fixed irradiation time of one hour. In the case of 1μm thick polymer film (**Figure 2a**) the DE signal increases until the SRG height of 100±10 nm (13min of irradiation at I=200 mW/cm$^2$) is reached. The inset of **Figure 2a** shows the *in-situ* recorded AFM micrograph during irradiation. The micrograph depicts the temporal evolution of the polymer topography with the vertical direction from top to bottom (red arrow at the bottom right corner) corresponding to the irradiation time. The AFM scanning starts in dark (flat topography), at the position marked by the dashed white line the irradiation with ±45° interference pattern (IP) is switched on. The distribution of the electric field vector relative to the topography maxima and minima is shown by the white arrows (**Figure 2a**). In the case of a 750nm thick polymer film the kinetics of the SRG growth is slower. The final SRG height is 160nm after one hour of irradiation (see also the inset in **Figure 2b,** showing the final *in-situ* AFM micrograph). The peak in the DE appears also at the SRG height of 100±10 nm (after 17min of irradiation). Further reduction of the film thickness to 500nm results in a drop of the final SRG height to 120nm (**Figure 2c**). In the DE, a saturation level at the characteristic height of 100±10 nm is reached. In the case of 200nm thick polymer film, the maximal SRG height after one hour of irradiation is 22nm (**Figure 2d**) and the DE signal does not show any drop.

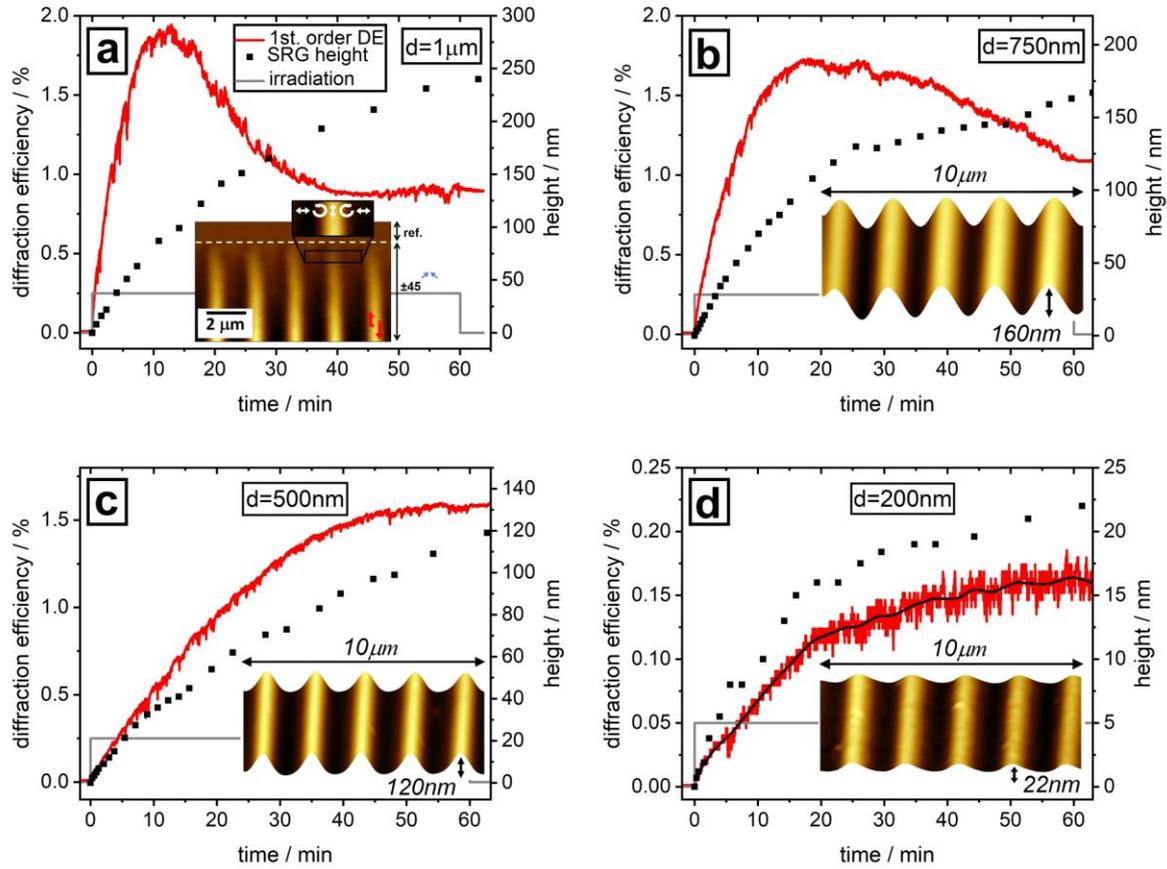

**Figure 2.** *In-situ* recorded SRG height (black dots) and diffraction efficiency (1st order DE, red curve) as a function of time during irradiation with $\pm 45°$ IP for different thicknesses of the Pazo polymer film: (a) 1μm, (b) 750nm, (c) 500nm and (d) 200nm. The inset in (a) shows the *in-situ* AFM micrograph of the polymer film deformation during irradiation. The direction of scanning (indicated by the red arrow at the bottom right corner) is from top to bottom showing the temporal evolution of film topography. The AFM scanning starts without irradiation (flat film), and at the position marked by the dashed white line, the irradiation ($\lambda = 491$nm, $I = 200$mW/cm$^2$, $\pm 45°$ configuration, $\Lambda = 2$μm) is switched on. The local polarization of the electric field relative to the topography maxima and minima is shown by white arrows. The insets in (b)-(d) illustrate the topographies of the final SRG measured by AFM.

In the following we explain that the distribution of the intensity within the diffraction spot is the result of the integration of light diffracted at different SRG modulation depths (different positions). Indeed, the grating does not have a constant modulation amplitude, *u*, over the inscribed area, but rather a circular shape set by the profile of the writing beams (see schematic representation in **Figure 4b**), due to the Gaussian profile of the writing beam. Since the spot sizes of the writing (4mm) and the probe beam (3mm) are comparable, the DE signal is acquired from the whole irradiated area where the probe beam is diffracted at spatially inhomogeneous modulated SRG amplitude.

In order to compare the data with a theoretical model, we have computed the diffracted light wave in the far field and the near field (see **Appendix** for details). The basic idea is to "shrink" the azo-polymer film into a phase grating that is observed in reflection. For our setup, the intensity of the diffraction spot in the *n*'th order is given by:

$$I_n(\boldsymbol{r}) = |R|^2 \, I_{\text{in}} \, |J_n(u(\boldsymbol{r}))|^2 \exp(-r^2/\sigma^2) \tag{1}$$

Here, $\boldsymbol{r}$ are two-dimensional coordinates in the plane perpendicular to the diffracted beam, $|R|^2 I_{\text{in}}$ is the total reflected intensity (typically only a few percent of $I_{\text{in}}$), $J_n$ is the n'th order Bessel function, the function $u(\boldsymbol{r})$ is proportional to the phase front modulation amplitude imprinted at position $\boldsymbol{r}$ by the film, and $\exp(-r^2/\sigma^2)$ gives the intensity profile of the probe beam. The diffraction efficiency $\eta_n$ is obtained by integrating Eq. (1) over $\boldsymbol{r}$ followed by normalization. The results of this calculation are illustrated in **Figure 3**. The angular profile of the diffraction spot is shown in **Figure 3a**, assuming that the probe beam diameter (σ) is similar in size to the SRG inscribed area ($w = \sigma$). As the grating amplitude grows (from bottom to top), a "dark ring" appears in the spot profile starting at black curve, $u = 3.5$, in **Figure 3a**).

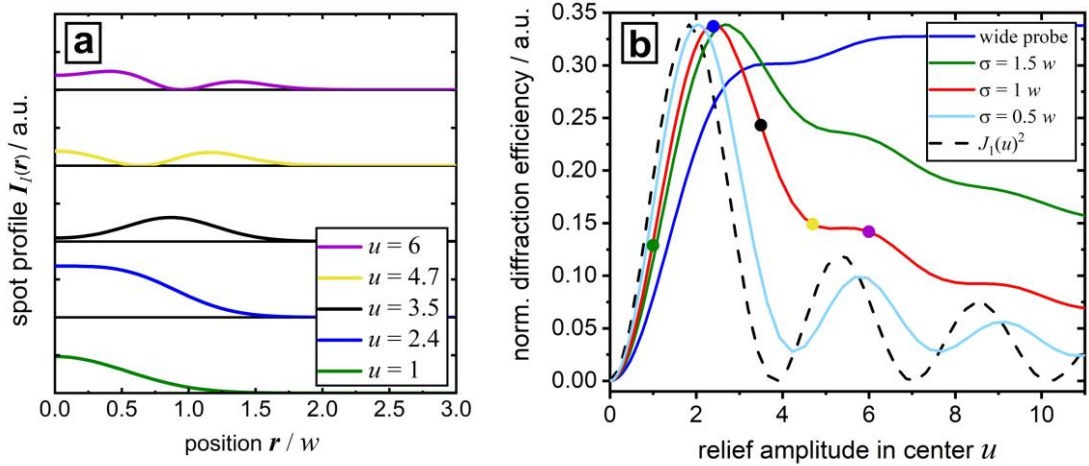

**Figure 3.** (a) Theoretical prediction for the spatial fine structure of the $n=1$ diffraction spot [Eq.(1)]. The parameter $u$ gives the modulation depth at the center of the phase grating, $w$ is the radius of the irradiated grating area (assuming a Gaussian profile). (b) Diffraction efficiency as a function of grating modulation $u$ (in the center), for different sizes σ of the probe beam. Solid lines: numerical calculation based on Eq. (A.14), integrating the diffracted intensity over the area of the $n=1$ spot. Dashed line: comparison to the Bessel function $J_1(u)^2$ evaluated at the grating center. All efficiencies are normalized to their maximal value. The dots mark the parameters chosen in (a).

This happens when the total diffracted intensity is beyond its maximum, as shown in **Figure 3b** (dots on the red curve). There, we also show the influence of the probe beam size: a narrow beam ($\sigma = 0.5w$) diffracts similar to a homogeneous grating, with an efficiency related to the Bessel function $|J_1|^2$ that oscillates beyond its first the maximum. For a much wider beam the diffraction efficiency increases monotonously.

We finally estimate the SRG height that corresponds to the first maximum of the 1$^{st.}$ order DE. In the reflection screen model, the optical path length permits to estimate the phase as $2\pi n' L/\lambda$ where $n'$ is the (real) refractive index at the probe wavelength. The probe beam passes from the substrate to the polymer-air interface and back. The relevant modulation depth of the path length $L$ is therefore twice the SRG height $\delta h$. This gives a phase modulation depth

$$u = \frac{4\pi n' \delta h}{\lambda} \qquad (2)$$

The maximum of the Bessel function $J_1$ appears at $u_{max} \approx 1.84$ so that we get a peak-to-bottom amplitude $2\,\delta h_{max} \approx 108$nm in reasonable agreement with the experimental value 100±10 nm ($n'_{633\text{nm}} = 1.7$). The agreement would become better (smaller value for $\delta h_{max}$) if we took into account a volume grating in the film which also contributes to the optical DE. Such an analysis will be presented in another paper.

We confirm experimentally the theoretical predictions shown in **Figure 3.** The spot size of the probe beam is reduced to 1mm and the DE signal, as well as the SRG amplitude, is recorded during irradiation at different position along the irradiated area (**Figure 4b**): at the center of the irradiated area (red curve in **Figure 4aI**), 1mm and 2mm away (red curves in **Figure 4aII** and **4aIII**). Here, the probe beam is still centered with respect to the AFM probe, but the center of the pump beam is shifted stepwise with respect to the AFM cantilever. As can be seen from **Figure 4a** with increasing the distance from the center, the height of the SRG decreases due to the Gaussian envelope of the writing beam intensity. The DE signal recorded in the center (I) has similar shape as in the case presented in **Figure 1**, but the extrema of curve (maxima and minima) are shifted to smaller SRG height as predicted in **Figure 3b**. At 1mm off center, the SRG height is smaller and the DE curve shifts to later times (red curve in **Figure 4aII**). The last measurement is performed 2mm away from the center at the edge of the illumination spot (see **Figure 4aIII**), here, the intensity is weak, and thus the maximal SRG height is 70nm, so that no drop in the DE became noticeable.

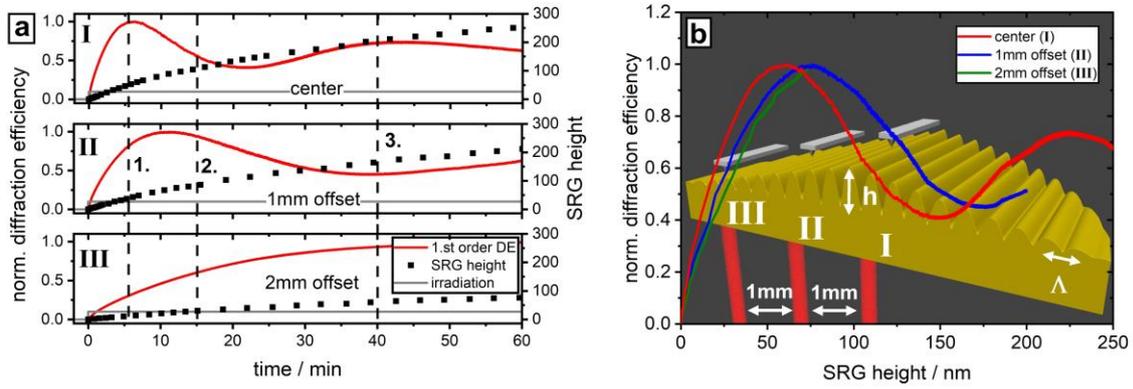

**Figure 4.** (a) *In-situ* recorded SRG height and normalized diffraction efficiency for the $\pm 45°$ interference pattern for different distances away from the center (center (I), 1mm (II) and 2mm (III) offset from center). The probe beam is focused to a four times smaller radius than the writing beam. The black dashed lines mark the moments where the fine structure of the 1$^{st}$ order diffracted beam is 1. Gaussian, 2. ring shaped (donut), and 3. ring shaped with a bright center. (b) Plot of DE vs. SRG amplitude (from (a)) measured at three different positions. A scheme is shown in the background for illustration. (Irradiation at λ=491nm with $I = 200$ mW/cm$^2$, $\Lambda = 2$ μm, $d_{Pazo} = 1$ μm)

**Conclusions**

We demonstrated that during continuous irradiation of a photosensitive polymer film with an interference pattern, the diffraction efficiency (DE) of a probe beam changes in a non-monotonous way, showing several maxima and minima, while the amplitude of the surface relief grating (SRG) is monotonously increasing. We used the Raman-Nath theory to compute the fine structure in the spatial profile of the diffraction spot that appears clearly when the DE goes through its maximum. The reason for this fine structure is that the spatial profile of the inscribed grating differs from an infinitely extended grating because of the Gaussian envelope of the irradiation pattern. Using a probe beam which matches the irradiated area in size, the profile of the diffraction spot changes as a function of the SRG height from a Gaussian through a hollow beam ("donut") to a ring structure with a bright center. For a narrow probe beam, the DE follows the theory for an infinite grating (squared *n*'th oder Bessel function), with a maximum 1st order DE appearing for SRG heights around ~100nm. In deeper reliefs, the DE is distinctly different from the Bessel function when the probe beam size is not much narrower than the irradiated area. These results are based on a simple reflecting phase screen model that combines the SRG and the birefringence grating in the bulk of the film into one complex reflection function.[39] These findings are confirmed in the experiment by probing the DE with a probe beam diameter much smaller than the inscribed area and aiming the probe at different positions of the SRG, while measuring the SRG amplitude simultaneously with an AFM. Only probe beams with sizes as narrow as half the waist of the writing beams ($\sigma = 0.5w$) diffract similar to an infinite grating with homogenous modulation depth. The Raman-Nath-based model we propose also comprehensively explains the fine structure within the diffraction spot.

ACKNOWLEDGMENTS: This research is supported by the Helmholtz Graduate School on Macromolecular Bioscience (Teltow, Germany). We thank Burkhard Stiller and Andreas Pucher for the fruitful discussions.

**Appendix**

We get a tractable calculation of the diffraction efficiency and the spot profile, when the sample is modelled as a reflecting phase screen. This means that the light wave, after (partial) reflection from the sample, is given by a complex reflection function that contains a spatially modulated phase[39]

$$\psi_{\text{out}}(\boldsymbol{r}) = R \exp[i\varphi(x,y)] \, \psi_{\text{in}}(\boldsymbol{r}) \tag{A.1}$$

where $R$ is the overall reflection amplitude and $x, y$ are coordinates parallel to the screen. (We write $R\,\psi_{\text{in}}(\boldsymbol{r})$ although this wave has actually a reversed propagation direction due to reflection from the non-modulated screen (specular order)). We first compute the diffraction pattern in the far field: it is given in the Fraunhofer-Kirchhoff approximation, by

$$I(\boldsymbol{q}) = |\tilde{\psi}_{\text{out}}(\boldsymbol{q})|^2 \quad (A.2)$$

where $\tilde{\psi}_{\text{out}}(\boldsymbol{q})$ is the the 2D Fourier transform of $\psi_{\text{out}}(\boldsymbol{r})$. We consider first the simple case of an infinitely extended phase grating where $\varphi(x,y) = u\sin(\boldsymbol{G}\cdot\boldsymbol{r})$ and $\boldsymbol{G}$ is the grating vector. Then, the outgoing wave is

$$\psi_{\text{out}}(\boldsymbol{r}) = R\exp[iu\sin(\boldsymbol{G}\cdot\boldsymbol{r})]\,\psi_{\text{in}}(\boldsymbol{r}) \quad (A.3)$$

The first factor (the phase modulation) is periodic in $\boldsymbol{r}$ along the direction $\boldsymbol{G}$ with period $\Lambda = 2\pi/G$. It can be expanded in a Fourier series

$$\exp[iu\sin(\boldsymbol{G}\cdot\boldsymbol{r})] = \sum_n J_n(u)\exp(in\boldsymbol{G}\cdot\boldsymbol{r}) \quad (A.4)$$

with coefficients $J_n(u)$ given by Bessel functions (Jacobi-Anger expansion). If the incoming wave is a plane wave, $\psi_{\text{in}}(\boldsymbol{r}) = \sqrt{I_{\text{in}}}\exp(i\boldsymbol{k}_{\text{in}}\cdot\boldsymbol{r})$, the outgoing wave contains the wave vectors

$$\boldsymbol{k}_n = \boldsymbol{k}_0 + n\boldsymbol{G} \quad (A.5)$$

We may choose the $x$-axis along $\boldsymbol{G}$ and introduce the diffraction angle $\theta_n$ via $k_{nx} = k\sin\theta_n$ with the wavenumber $k = 2\pi/\lambda$. The discrete diffraction orders appear under the angles (Bragg-Laue condition)

$$\sin\theta_n - \sin\theta_0 = \frac{nG\lambda}{2\pi} \quad (A.6)$$

where $\lambda$ is the wavelength of the wave. For an infinite grating, the diffraction efficiencies are thus given by the squared coefficients in **Eq.** (A.4):

$$\eta_n = \frac{I_n}{I_\text{in}} = |R|^2 |J_n(u)|^2 \qquad (A.7)$$

In the experiment, the grating does not have a constant modulation amplitude $u$, but rather a circular shape set by the profile of the writing beams. We assume here a Gaussian grating profile

$$\varphi(x,y) = u(\rho) \sin Gx \qquad u(\rho) = u\, e^{-\rho^2/2w^2} \qquad (A.8)$$

with $\rho^2 = x^2 + y^2$. The width $w$ is much larger than the grating period $\Lambda$. This motivates an approximate evaluation of the Fourier transform $\tilde{\psi}_\text{out}(\boldsymbol{q})$ (the far field)

$$\tilde{\psi}_\text{out}(\boldsymbol{q}) = R\sqrt{I_\text{in}} \int dx\, dy\, e^{iu(\rho)\sin Gx} \exp[i(\boldsymbol{k}_\text{in} - \boldsymbol{q}) \cdot \boldsymbol{r}] \qquad (A.9)$$

that we explain in the following. We split the integral along $x$ into periods of size $\Lambda$ centered in $x_l = l\Lambda$ and replace $x \to x_l + x$ with $-\Lambda/2 \le x \le \Lambda/2$ and $l = 0, \pm 1, \pm 2, \cdots$. We also focus on a far-field direction close to the $n$'th order and set $q_x = k_{\text{in }x} + nG + \delta q$ with $\delta q \ll G$ (angular fine structure of the diffraction spot). Observe that $(q_x - k_{\text{in }x})(x_l + x) \equiv \delta q\, x_l + (nG + \delta q)x \bmod 2\pi$. We apply a multiple-scale approximation and assume that the grating modulation $u(\rho)$ varies slowly across the grating period $\Lambda$. This means that we can set

$$u(\rho) \approx u(\rho_l), \qquad \rho_l^2 = x_l^2 + y^2 \qquad (A.10)$$

The integral along x over one grating period then gives

$$\int_{-\Lambda/2}^{\Lambda/2} dx\, e^{iu(\rho_l)\sin Gx} \exp[-i(nG + \delta q)x] \approx \Lambda J_n(u(\rho_l)) \qquad (A.11)$$

where a small phase ($|x\, \delta q| \le \Lambda \delta q$) was neglected. (This will be justified from the end result **Eq.** (A.13) below.) The sum over the grating periods is replaced back by an integral

$$\tilde{\psi}_{\text{out}}(\boldsymbol{q}) \approx R\sqrt{I_{\text{in}}} \sum_l \Lambda \int dy\, J_n\big(u(\rho_l)\big)\, e^{-i\delta\boldsymbol{q}\cdot\boldsymbol{r}_l}$$

$$\approx R\sqrt{I_{\text{in}}} \int d^2r\, J_n\big(u(\rho)\big)\, e^{-i\delta\boldsymbol{q}\cdot\boldsymbol{r}} \tag{A.12}$$

This is the key result of the calculation: the shape of the diffracted beam (as measured by the angular deviation $\delta q$ from the $n$'th order) is the Fourier transform of the *local* diffraction amplitude (the Bessel function $J_n(u(\rho))$).

We finally compute the near-field profile of the diffraction spot. It is easy to read off the back Fourier transform from **Eq.** (A.12). We get for the beam shape in the $n$'th order:

$$\psi_n(\boldsymbol{r}) = R\sqrt{I_{\text{in}}}\, J_n\big(u(\rho)\big) \tag{A.13}$$

This proves Eq. (1) in the main text.

To evaluate Eq. (A.12), one simplification is possible by switching to polar coordinates: $\exp(-i\delta\boldsymbol{q}\cdot\boldsymbol{r}) = \exp(-i\rho\,\delta q \cos\phi)$. The $\phi$ integral gives $2\pi J_0(\rho\,\delta q)$, and we finally get

$$\tilde{\psi}_{\text{out}}(\boldsymbol{q}) \approx 2\pi R\sqrt{I_{\text{in}}} \int_0^\infty \rho\, d\rho\, J_n\big(u\, e^{-\rho^2/2w^2}\big)\, J_0(\rho\,\delta q) \tag{A.14}$$

For large $\rho \gg w$, the Bessel function $J_n$ becomes proportional to $e^{-n\rho^2/2w^2}$ so that the integral converges for $n > 0$. Its numerical evaluation presents no difficulties. We can estimate its typical width as a function of the scattering wave vector as $\delta q \sim 1/w$. Since we assumed a scale $w$ for the grating profile much wider than the period $\Lambda$, the phase neglected in Eq. (A.11) is at most $\mathcal{O}(\Lambda/w) \ll 1$.

Finally, we consider the case that incident and specular beams have a Gaussian profile

$$\psi_{\text{in}}(\boldsymbol{r}) = R\sqrt{I_{\text{in}}}e^{i\boldsymbol{k}_{\text{in}}\cdot\boldsymbol{r}}\exp(-\rho^2/2\sigma^2) \tag{A.15}$$

We assume that the waist of the probe beam $\sigma$ is also larger than the grating period. The preceding calculation can be carried through, and we get Eq. (A.14) with an additional factor $\exp(-\rho^2/2\sigma^2)$ under the integral. The predictions of this theory are illustrated in **Figure 3**. The angular profile of the diffraction spot is shown in **Figure 3a**, assuming that the probe beam is matched in size to the grating covered area ($\sigma = w$). As the grating amplitude grows (from bottom to top), a "dark ring" enters the spot profile (top curves). This happens when the total diffracted intensity is beyond its maximum, as shown in **Figure 3b** (dots on the red curve). There, we also show the influence of the probe beam size: a narrow beam ($\sigma = w/2$) diffracts similar to an infinite grating, with an efficiency related to the Bessel function $|J_1(u)|^2$ that oscillates beyond the maximum efficiency. For a much wider beam ($\sigma \gg w$), the diffraction efficiency increases monotonously. An analytical proof of this property is given in the Supporting Information.

**Graphical Abstract**

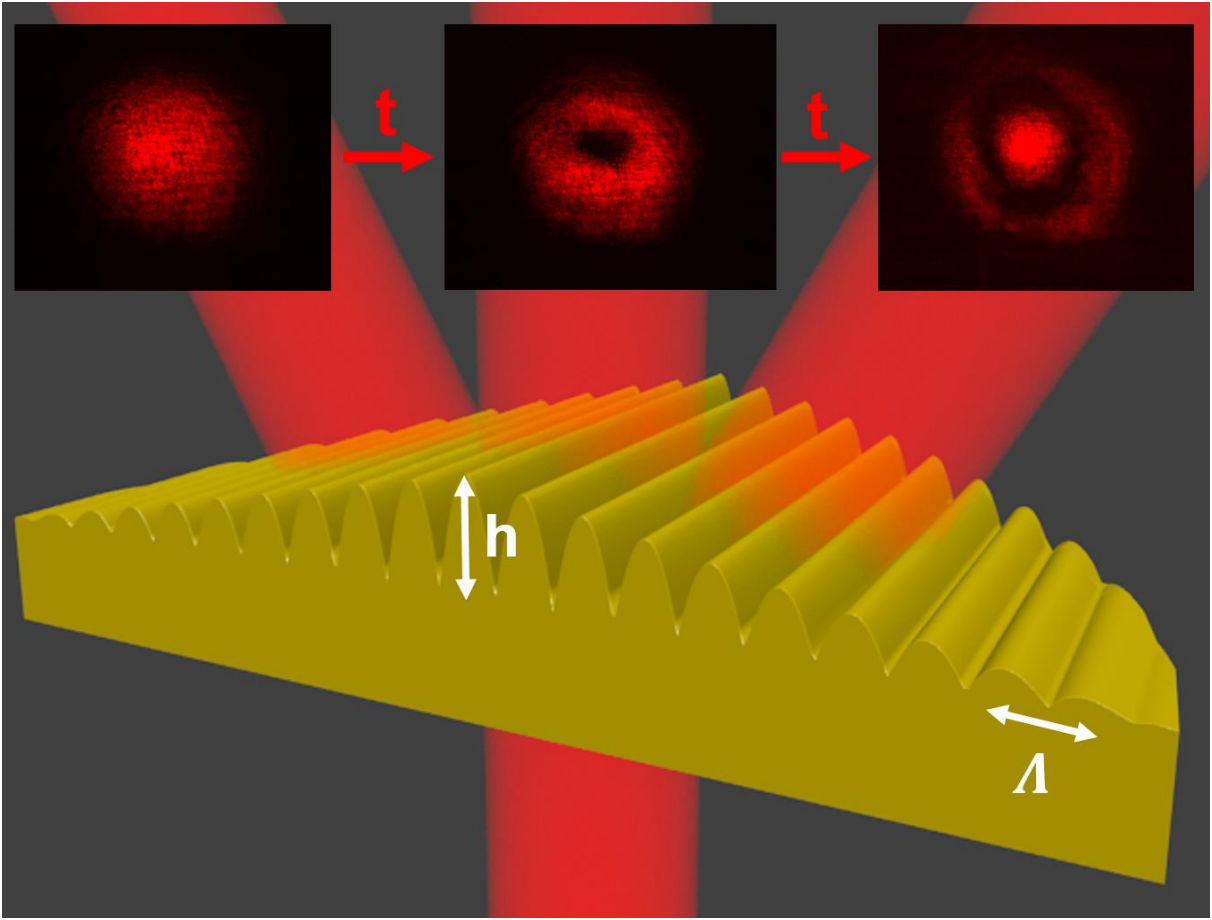

# Supporting Information

# Solving an old puzzle: Fine structure of diffraction spots from an azo-polymer surface relief grating

Joachim Jelken, Carsten Henkel, Svetlana Santer

*Institute of Physics and Astronomy, University of Potsdam, 14476 Potsdam, Germany*

AUTHOR EMAIL ADDRESS: henkel@uni-potsdam.de

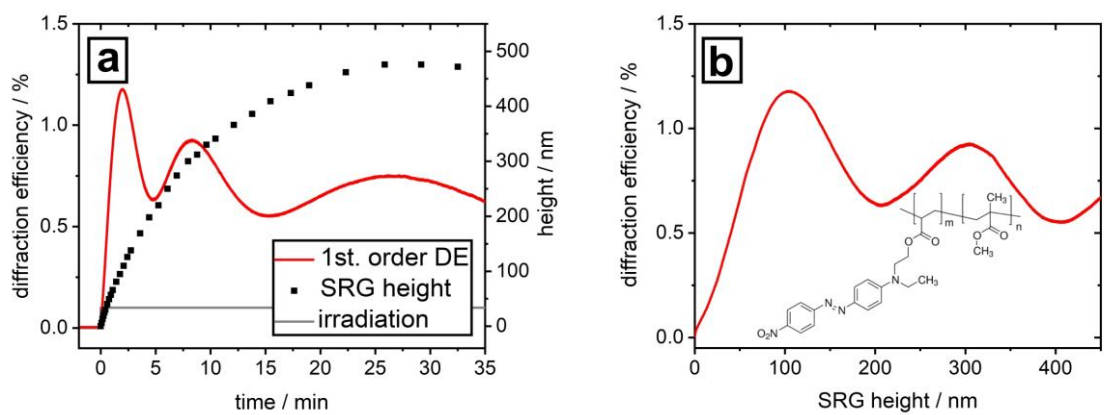

**Figure S1.** (a) *In-situ* recorded SRG height and $1^{st.}$ order diffraction efficiency (DE) of the poly(MMA-co-DR1A) polymer film during irradiation with a $\pm 45°$ interference pattern. (b) Dependence of the DE signal on SRG height. The chemical structure of the polymer is shown in the inset. ($\lambda = 491$nm, $I = 200$mW/cm$^2$, $\Lambda = 2$μm, $d = 600$nm)

**video file name (separate file, not embedded in this document):**

**Figure_S2_change_in_spatial_profile.mp4**

**Figure S2.** Video (accelerated) showing the spatial fine structure of the 1$^{st}$ order diffraction spot as a function of time while irradiation with a $\pm 45°$ interference pattern (IP) of $\lambda = 491$nm light ($I = 200$mW/cm$^2$, $\Lambda = 2$µm grating period, Pazo film, thickness 1µm). The spatial profile is changing with time from a Gaussian to a ring-shaped intensity distribution ("donut") and finally to a ring structure with a bright center.

We derive here a property mentioned in the Appendix, namely that the diffraction efficiency increases monotonously with grating amplitude for a wide probe beam. We compute the integral of the diffracted intensity over the spot profile (the $q$-integral is restricted to the $n$th diffraction order)

$$I_n = \int \frac{d^2q}{(2\pi)^2} |\tilde{\psi}_{out}(\boldsymbol{q})|^2 \tag{A.16}$$

Inserting Eq. (A.12) (with $\sigma = \infty$) and performing the $\boldsymbol{q}$-integral first, we get (Parseval-Plancherel formula)

$$I_n = I_{in}|R|^2 \int d^2r\, J_n(u(\rho))^2 \tag{A.17}$$

Using $u(\rho) = u\, e^{-\rho^2/2w^2}$ of Eq.(A.8), the derivative with respect to the modulation amplitude $u$ in the center of the grating is

$$\frac{\partial I_n}{\partial u} = 2I_{in}|R|^2 \int d^2r\, e^{-\rho^2/2w^2} J_n'(u(\rho))\, J_n(u(\rho)) \tag{A.18}$$

where $J_n'$ is the derivative of the Bessel function. We now show that this quantity is positive. In polar coordinates, the angular integral is trivial, and the substitution $z = u(\rho)$ permits to evaluate the radial integral in closed form:

$$\frac{\partial I_n}{\partial u} = \frac{4\pi}{u} I_{in}|R|^2 \int_0^\infty \rho d\rho\, u(\rho) J_n'(u(\rho))\, J_n(u(\rho))$$

$$= \frac{4\pi w^2}{u} I_{in}|R|^2 \int_0^u dz\, J_n'(z)\, J_n(z)$$

$$= \frac{2\pi w^2}{u} I_{in}|R|^2 J_n(u)^2 \geq 0 \tag{A.19}$$

which is always a positive quantity.